\documentclass[twocolumn,showpacs,preprintnumbers,amsmath,amssymb]{revtex4}
\usepackage{graphicx}
\usepackage{dcolumn}
\usepackage{bm}

\renewcommand{\vec}[1]{\mbox{\boldmath $#1$}}
\newcommand{\sgn}{\mbox{sgn}}
\newcommand{\erf}{\mbox{erf}}

\newcommand{\epsfig}[2]{
\begin{center}
\scalebox{0.37}{\includegraphics{#1}} 
\end{center}
}

\begin{document}
\preprint{APS/123-QED}
\title{Synapse efficiency diverges due to synaptic pruning following
over-growth}
\author{Kazushi Mimura${}^{\dagger}$}
\email{mimura@kobe-kosen.ac.jp}
\author{Tomoyuki Kimoto${}^{\ddagger}$}
\author{Masato  Okada${}^{\S}$}
\affiliation{
${}^{\dagger}$
Department of Electrical Engineering, 
Kobe City College of Technology, 
Gakuenhigashi-machi 8-3, Nishi-ku, Kobe, Hyogo, 651-2194 Japan\\
${}^{\ddagger}$
Oita National College of Technology, 
Maki 1666, Oita-shi, Oita, 870-0152 Japan\\
${}^{\S}$
Laboratory for Mathematical Neuroscience,
RIKEN Brain Science Institute, Saitama 351-0198, Japan, \\
"Intelligent Cooperation and Control", PRESTO, JST, 
c/o RIKEN BSI, Saitama 351-0198, Japan, \\
ERATO Kawato Dynamic Brain Project, 
2-2 Hikaridai, Seika-cho, Soraku-gun, Kyoto 619-0288, Japan
}
\date{\today}
\begin{abstract}
\label{sec.sparse_noise.abst}
In the development of the brain, 
it is known that synapses are pruned following
over-growth. 
This pruning following over-growth seems to be a universal phenomenon 
that occurs in almost all areas -- visual cortex, 
motor area, association area, and so on. 
It has been shown numerically 
that the synapse efficiency is increased by systematic deletion. 
We discuss the synapse efficiency 
to evaluate the effect of pruning following over-growth, and 
analytically show that 
the synapse efficiency diverges as $O(|\log c|)$ 
at the limit where connecting rate $c$ is extremely small. 
Under a fixed synapse number criterion, 
the optimal connecting rate, which maximize memory performance, exists. 
\end{abstract}
\maketitle


\section{Introduction}
\label{sec.sparse_noise.intro}

In this paper, we analytically discuss synapse efficiency 
to evaluate effects of pruning following over-growth 
during brain development, 
within the framework of auto-correlation-type associative memory. 

Because this pruning following over-growth seems to be a universal phenomenon 
that occurs in almost all areas 
-- visual cortex, motor area, association area, and so on 
\cite{
Huttenlocker1979,Huttenlocker1982,
Bourgeois1993,Takacs1986,Innocenti1995,Eckehoff1991,
Rakic1994,Stryker1986,Sur1990,Wolff1995} 
-- 
we discuss the meaning of its function from a universal viewpoint 
rather than in terms of particular properties in each area. 
Of course, to discuss this phenomenon 
as a universal property of a neural network model, 
we need to choose an appropriate model. 

Artificial neural network models are roughly classified into two types: 
feed forward models and recurrent models. 
Various learning rules are applied to the architectures of these models, 
and correlation learning corresponding to the Hebb rule can be considered 
a prototype of any other learning rules. 
For instance, correlation learning can be regarded 
as a first-order approximation of the orthogonal projection matrix, 
because the orthogonal projection matrix 
can be expanded by correlation matrices \cite{Yanai1996}. 
In this respect, we can naturally regard 
a correlation-type associative memory model 
as one prototype of the neural network models of the brain. 
For example, Amit et al. discussed 
the function of the column of anterior ventral temporal cortex 
by means of a model based on correlation-type associative memory model 
\cite{Amit1994,Griniasty1993}. 
Also, Sompolinsky discussed the effect of dilution. 
He assumed that the capacity is proportional 
to the number of remaining bonds, 
and pointed out that 
a synapse efficiency of diluted network, 
which is storage capacity per a connecting rate, 
is higher than a full connected network's one \cite{Sompolinsky1986b}. 
Chechik et al. discussed the significance 
of the function of the pruning following over-growth 
on the basis of a correlation-type associative memory model 
\cite{Chechik1998}. 
They pointed out that 
a memory performance, which is stored pattern number per synapse number, 
is maximized by systematic deletion 
that cuts synapses that are lightly weighted. 
However, while it is qualitatively obvious 
that synapse efficiency and memory performance 
are increased by a systematic deletion, 
we also need to consider the increase of synapse efficiency quantitatively. 

In this paper, 
we quantitatively compare 
the effectiveness of systematic deletion to that with random deletion 
on the basis of an auto-correlation-type associative memory model. 
In this model, 
one neuron is connected to other neurons 
with a proportion of $c$, 
where $c$ is called the connecting rate. 
Systematic deletion is considered as a kind of nonlinear correlation learning 
\cite{Okada1998}. 
At the limit where the number of neurons $N$ is extremely large, 
it is known that 
random deletion and nonlinear correlation learning 
can be transformed into correlation learning with synaptic noise
\cite{Sompolinsky1986b,Okada1998}. 
These two types of deletion, systematic and random, are strongly related 
to multiplicative synaptic noise. 
First, we investigated 
the dependence of storage capacity on multiplicative synaptic noise. 
At the limit where multiplicative synaptic noise is extremely large, 
we show that storage capacity is inversely proportional 
to the variance of the multiplicative synaptic noise. 
From this result, 
we analytically derive that 
the synapse efficiency in the case of systematic deletion 
diverges as $O(|\log c|)$
at the limit where the connecting rate $c$ is extremely small. 
We also show that the synapse efficiency 
in the case of systematic deletion becomes $2 |\log c|$ times 
as large as that of random deletion. 

In addition to such the fixed neuron number criterion 
as the synapse efficiency, 
a fixed synapse number criterion could be discussed. 
At the synaptic growth stage, 
it is natural to assume 
that metabolic energy resources are restricted. 
When metabolic energy resources are limited, 
it is also important 
that the effect of synaptic pruning is discussed 
under limited synapse number. 
Under this criterion, 
the optimal connecting rate, which maximize memory performance, exists.
These optimal connecting rates are in agreement 
with the computer simulation results 
given by Chechik et al \cite{Chechik1998}.


\section{Model}
\label{sec.sparse_noise.model}

Sompolinsky  discussed 
the effects of synaptic noise and nonlinear synapse 
by means of the replica method \cite{Sompolinsky1986b}. 
However, 
symmetry of the synaptic connections $J_{ij}=J_{ji}$ 
is required in the replica method 
since the existence of the Ljapunov function is necessary. 
Therefore, there was a problem that 
the symmetry regarding synaptic noise 
had to be assumed in the Sompolinsky theory. 
To avoid this problem, Okada et al. discussed 
additive synaptic noise, multiplicative synaptic noise, 
random synaptic deletion, and nonlinear synapse 
by means of the self-consistent signal-to-noise analysis (SCSNA) 
\cite{Okada1998}. 
They showed that 
additive synaptic noise, 
random synaptic deletion, nonlinear synapse 
can be transformed into multiplicative synaptic noise. 

Here, 
we discuss the synchronous dynamics as, 
\begin{equation}
x_i = F(\sum_{j \neq i}^N J_{ij} x_j + h), 
\label{5eq:sparse.equilibrium}
\end{equation}
where $F$ is the response function, 
$x_i$ is the output activity of neuron $i$, 
and $-h$ is the threshold of each neuron. 
Every component $\xi_i^\mu$ in a memorized pattern 
$\mbox{\boldmath $\xi$}^\mu$ is an independent random variable, 
\begin{equation}
{\rm Prob}[\xi_i^\mu=\pm 1]=\frac{1 \pm a}2,
\label{5eq:sparse.pattern}
\end{equation}
and the generated patterns are called sparse pattern with bias $a \; (-1<a<1)$. 
We have determined that the firing rate of states in the retrieval phase 
is the same for each memorized pattern \cite{Amit1987,Amari1989}. 
In this case, threshold $-h$ can be determined as, 
\begin{equation}
a = \frac{1}{N} \sum_{i=1}^N \mbox{sgn}(\sum_{j \neq i } J_{ij} x_i + h).
\label{5eq:sparse.constraint.1}
\end{equation}
The firing rate becomes $f=(1+a)/2$ at the bias $a$. 

Additive synaptic noise, multiplicative synaptic noise, 
random synaptic deletion, and nonlinear synapse can be introduced 
by synaptic connections in the following manner. 

In the case of additive synaptic noise, 
synaptic connections are constituted as, 
\begin{equation}
J_{ij} = \frac{J}{N(1-a^2)}
  \sum_{\mu = 1}^{\alpha N} (\xi_i^\mu-a) (\xi_j^\mu-a)
  + \delta_{ij}, 
\label{5eq:add.noise.J}
\end{equation}
where $\delta_{ij}$ is the additive synaptic noise. 
The symmetric additive synaptic noise $\delta_{ij}$ and $\delta_{ji}$ 
are generated according to the probability, 
\begin{equation}
\delta_{ij} \sim N (0,\frac{\delta_A^2}{N} ), \quad 
\delta_{ij}=\delta_{ji}, 
\end{equation}
where $\delta_A^2$ is the absolute strength 
of the additive synaptic noise. 
The parameter $\delta_A$ is assumed to be $O(1)$. 
This means that the synaptic connection $J_{ij}$ is $O(1/\sqrt{N})$. 
It is useful to define the parameter $\Delta_A$ as 
\begin{equation}
\Delta_A \equiv \frac{\delta_A}{J/(1-a^2)}, 
\end{equation}
which measures the relative strength of the noise 
and we call the parameter $\Delta_A^2$ 
the variance of the additive synaptic noise. 
Therefore, we define the probability 
to generate the additive synaptic noise $\delta_{ij}$ as 
\begin{equation}
\delta_{ij} \sim N (0,\frac{J^2}{N(1-a^2)^2} \Delta_A^2 ), \quad 
\delta_{ij}=\delta_{ji}. 
\label{5eq:add.noise.pdf}
\end{equation}

In the case of multiplicative synaptic noise, 
synaptic connections are constituted as, 
\begin{equation}
J_{ij} 
= \frac{1+\varepsilon_{ij}}{N(1-a^2)}
  \sum_{\mu = 1}^{\alpha N} (\xi_i^\mu-a) (\xi_j^\mu-a), 
\label{5eq:mul.noise.J}
\end{equation}
where $\varepsilon_{ij}$ is multiplicative synaptic noise. 
The symmetric multiplicative noise 
$\varepsilon_{ij}$ and $\varepsilon_{ji}$ 
are generated according to the probability, 
\begin{equation}
\varepsilon_{ij} \sim N(0,\Delta_M^2), \quad
\varepsilon_{ij}=\varepsilon_{ji}, 
\label{5eq:mul.noise.pdf}
\end{equation}
where $\Delta_M^2$ is the variance of the multiplicative synaptic noise. 

In the model of random synaptic deletion, 
synaptic connections are constituted as, 
\begin{equation}
J_{ij} = \frac{c_{ij}}{Nc(1-a^2)}
\sum_{\mu = 1}^{\alpha N} (\xi_i^\mu-a) (\xi_j^\mu-a), 
\label{5eq:random.cut.J}
\end{equation}
where $c_{ij}$ is a cut coefficient. 
The synapse that is cut is represented 
by the cut coefficient $c_{ij}=0$. 
In the case of symmetric random deletion, 
the cut coefficients $c_{ij}$ and $c_{ji}$ 
are generated according to the probability, 
\begin{equation}
\mbox{Prob}[c_{ij}=1] = 1-\mbox{Prob}[c_{ij}=0]=c, \quad 
c_{ij}=c_{ji}, 
\label{5eq:cut.pdf}
\end{equation}
where $c$ is the connecting rate. 

In the model of nonlinear synapse, 
synaptic connections are constituted as, 
\begin{eqnarray}
J_{ij} &=& \frac{\sqrt{p}}{N} f(T_{ij}), 
\label{5eq:nonlinear.learning} \\
T_{ij} &=& \frac{1}{\sqrt{p}(1-a^2)} \sum_{\mu = 1}^{p} 
(\xi_{i}^{\mu} - a) (\xi_{j}^{\mu} - a) \nonumber \\
& \sim & N(0,1) , 
\label{5eq:nonlinear.T}
\end{eqnarray}
where $p=\alpha N$. 
The nonlinear synapse is introduced 
by applying the nonlinear function $f(x)$ 
to the conventional Hebbian connection matrix $T_{ij}$. 


\subsection{Systematic deletion by nonlinear synapse}
\label{sec.sparse_noise.systematic.cut}

\begin{figure}
\epsfig{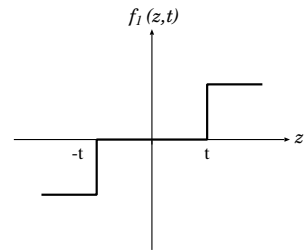}{6cm}
\caption[Clipped synapse]{Clipped synapse.}
\label{5fig:nonlinear.clip}
\end{figure}

\begin{figure}
\epsfig{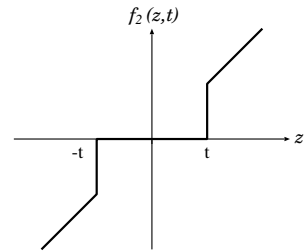}{6cm}
\caption[Minimal value deletion]{Minimal value deletion.}
\label{5fig:nonlinear.minimal}
\end{figure}

\begin{figure}
\epsfig{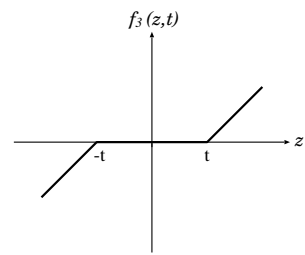}{6cm}
\caption[Compress deletion]{Compressed deletion.}
\label{5fig:nonlinear.compress}
\end{figure}

Chechik et al. pointed out 
that a memory performance, 
which is a storage pattern number per a synapse number, is maximized 
by systematic deletion that cuts synapses that are lightly weighted
 \cite{Chechik1998,Amit1994,Griniasty1993}. 
Such a systematic deletion can be represented 
by the nonlinear function $f(x)$ for a nonlinear synapse. 
In accordance with Chechik et al.,
we discuss three types of nonlinear functions 
(Figs.\ref{5fig:nonlinear.clip}, 
\ref{5fig:nonlinear.minimal}, and \ref{5fig:nonlinear.compress}). 
Fig.\ref{5fig:nonlinear.clip} shows clipped modification that is 
discussed generally as 
\begin{equation}
f_1(z,t) = \left\{ \begin{array}{ll} 
                \sgn(z),  & |z|>t  \\
                0,        & \mbox{otherwise} .
           \end{array}
           \right.
\label{5eq:nonlinear.clip}
\end{equation}

Chechik et al. also obtained 
the nonlinear functions shown 
in Figs.\ref{5fig:nonlinear.minimal} and \ref{5fig:nonlinear.compress} 
by applying the following optimization principles 
\cite{Chechik1998,Meilijson1996}. 
In order to evaluate the effect of synaptic pruning on the network's retrieval performance, 
Chechik et al. study its effect on the signal-to-noise ratio (S/N) of 
the internal field $h_i \equiv \sum_{j \ne i} J_{ij} x_j + h$ \cite{Chechik1998}. 
The S/N is calculated by analyzing the moments of the internal field and was given as 
\begin{eqnarray}
S/N 
&=& \frac{E[h_i |\xi_i^\mu=1]-E[h_i |\xi_i^\mu=-1]} {\sqrt{V[h_i|\xi_i^\mu]}} \nonumber \\
&\propto& \frac{E[zf(z)]}{\sqrt{E[f(z)^2]}} 
= \frac{\displaystyle{\int_{-\infty}^\infty Dz \; zf(z)}}
    {\displaystyle{\int_{-\infty}^\infty Dz \; f(z)^2}} \nonumber \\
&\equiv& \rho(f(z),z), 
\end{eqnarray}
where $z$ has standard normal distribution, i.e., $E[z^2]=V[z]=1$ and $Dz$ is Gaussian measure defined as 
\begin{equation}
Dz = \frac {dz}{\sqrt {2 \pi}} \exp (-\frac{z^2}2), 
\end{equation}
and $E,V$ denote the operators to calculate the expectation and the variance 
for the random variables $\xi_i^\mu \; (i=1,\cdots ,N, \mu=1, \cdots ,\alpha N)$,
respectively. 
The function $\rho(f(z),z)$ denotes the correlation coefficient. 
Chechik et al. considered the piecewise linear function like following nonlinear functions. 
In order to find the best nonlinear function $f(z)$, 
we should maximize $\rho(f(z),z)$, which is invariant to scaling. 
Namely, the best nonlinear function $f(x)$ is obtained 
by maximizing $E[zf(z)]$ under the condition that $E[f(z)^2]$ is constant. 
Let $\gamma$ be the Lagrange multiplier, it is sufficient to solve 
\begin{equation}
\int_{-\infty}^\infty Dz \; zf(z)
- \gamma \left( \int_{-\infty}^\infty Dz \; f(z)^2 -c_0 \right)
\to \mbox{max}, 
\end{equation}
for some constant $c_0$. 
Since the synaptic connection before acting the nonlinear function $T_{ij}$ obeys 
a Gaussian distribution $N(0,1)$, Eq.(\ref{5eq:nonlinear.T}) is averaged over all of the synaptic connections. 


Thus, the nonlinear function shown in Fig.\ref{5fig:nonlinear.minimal}, 
\begin{equation}
f_2(z,t) = \left\{ \begin{array}{ll} 
                z,  & |z|>t \\
                0,  & \mbox{otherwise} ,
           \end{array}
           \right.
\label{5eq:nonlinear.minimal}
\end{equation}
is also obtained. 
The deletion by this nonlinear function is called minimal value deletion. 
Similarly, by adding the condition 
that the total strength of synaptic connection 
$\int Dz \; |f(z)|$ is constant, 
the nonlinear function 
\begin{eqnarray}
f_3(z,t) &=& \left\{ \begin{array}{ll} 
                z-\sgn(z)t,  & |z|>t  \\
                0,          & \mbox{otherwise} 
             \end{array}
             \right. \nonumber \\
         &=& f_2(z,t)-tf_1(z,t), 
\label{5eq:nonlinear.compress}
\end{eqnarray}
is obtained. 
The deletion by this nonlinear function is called compressed deletion. 
We discuss systematic deletion 
by using these three types of nonlinear functions 
$f_1(z,t),f_2(z,t)$ and $f_3(z,t)$ given by Chechik et al \cite{Chechik1998}. 

\section{Results}
\label{sec.sparse_noise.result}

In this section, 
the results concerning the multiplicative synaptic noise, 
the random deletion, and the nonlinear synapse are shown, at the limit 
where the effect of these deletions is extremely large. 

The SCSNA starts from the fixed-point equations 
for the dynamics of an $N$-neuron network 
as Eq.(\ref{5eq:sparse.equilibrium}). 
The results of the SCSNA for the symmetric additive synaptic noise 
are summarized by the following order-parameter equations 
(see Appendix \ref{appendix.scsna}) : 
\begin{eqnarray}
m &=& \frac{1}{1-a^2} \int Dz < (\xi -a) Y(z;\xi ) >_\xi , 
\label{eq:order_parameter_equation_m} \\
q &=& \int Dz < Y(z;\xi )^2 >_\xi , 
\label{eq:order_parameter_equation_q} \\
U &=& \frac{1}{\sigma} \int Dz \; z < Y(z;\xi ) >_\xi , 
\label{eq:order_parameter_equation_U} \\
\sigma^2 &=& \frac{\alpha J^2 q}{(1-JU)^2} 
+ \frac{J^2}{(1-a^2)^2} \Delta_A^2 q. 
\label{eq:order_parameter_equation_sigma} \\
Y(z;\xi ) 
&=& F \biggl( J (\xi-a) m + \sigma z + h \nonumber \\
& & + \biggl[ \frac{\alpha J^2 U}{1-JU}+\frac {J^2}{(1-a^2)^2} \Delta_A^2 \biggr] 
Y(z;\xi ) \biggr) , \label{eq:effevtive_response_function}
\end{eqnarray}
where $< \cdots >_\xi$ implies averaging over the target pattern, 
$m$ is the overlap between the 1st memory pattern $\vec{\xi}^1$ 
and the equilibrium state $\vec{x}$ is defined as 
\begin{equation}
m = \frac{1}{N(1-a^2)} \sum_{i=1}^N (\xi^1_i-a) x_i, 
\label{eq:definition_overlap}
\end{equation}
note that generality is kept 
even if the overlap was defined by only the 1st memory pattern,
$q$ is Edwards-Anderson order-parameter, 
$U$ is a kind of the susceptibility, which measures 
sensitivity of neuron output with respect to the external input, 
$Y(z;\xi )$ is effective response function, 
$\sigma^2$ is the variance of the noise. 
We set the output function $F(x)={\rm sgn}(x)-a$ in the following sections, 
where domain of the $x$ variable is $F(x)=1-a$ when $x \le 0$, otherwise $F(x)=-1-a$.

According to Okada et al. \cite{Okada1998}, 
the symmetric additive synaptic noise, the symmetric random deletion, 
and the nonlinear synapse can be transformed 
into the symmetric multiplicative synaptic noise as follows 
(see Appendix \ref{appendix.equivalent_noise}): 
the additive synaptic noise is 
\begin{equation}
\Delta_M^2 = \frac{\Delta_A^2}{\alpha(1-a^2)^2},
\label{5eq:add.to.mul}
\end{equation}
the random deletion is 
\begin{equation}
\Delta_M^2 = \frac {1-c}{c},
\label{5eq:cut.to.mul}
\end{equation}
and the nonlinear synapse is 
\begin{equation}
\Delta_M^2 = \frac{{\tilde{J}}^2}{J^2} -1,
\label{5eq:nonlinear.to.mul}
\end{equation}
where $J, \tilde{J^2}$ are 
\begin{eqnarray}
J &=& \int Dz \; z f(z),
\label{5eq:nonlinear.J} \\
{\tilde J}^2 &=& \int Dz \; f(z)^2. 
\label{5eq:nonlinear.tildeJ}
\end{eqnarray}
In the following sections, symmetries of the additive synaptic noise, 
the multiplicative synaptic noise and random deletion are assumed.　
Storage capacity can be obtained 
by solving the order-parameter equations. 

\begin{figure}
\epsfig{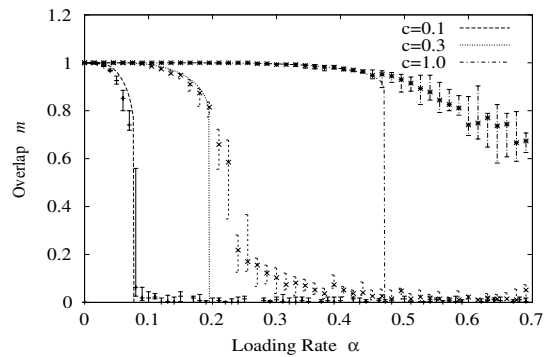}{8cm}
\caption{
Overlaps in the random deletion network. 
The curves represent the theoritical results. 
The dots represent simulation results with 
$N=3000$ and $f=0.1$ 
for the connecting rate $c=0.1$,$c=0.3$, and $c=1.0$. 
}
\label{5fig:computer_simulation}
\end{figure}

Figure \ref{5fig:computer_simulation} shows 
$m(\alpha )$ curves in the random deletion network 
with the number of neurons $N=3000$ and the firing rate $f=0.1$ 
for the connecting rate $c=0.1$,$c=0.3$, and $c=1.0$. 
It can be confirmed that 
the theoretical results of the SCSNA are in good agreement 
with the computer simulation results from Fig.\ref{5fig:computer_simulation}. 
Since it is known that 
theoretical results obtained by means of the SCSNA 
are generally in good agreement 
with the results obtained through the computer simulations 
using various models that include synaptic noise, 
we treat the results by means of the SCSNA only
\cite{Shiino1992,Shiino1993,Okada1995,Okada1996,Okada1998,Mimura1998-1,Mimura1998-2,Kimoto2001}. 

Through the relationships 
of Eqs.(\ref{5eq:add.to.mul})-(\ref{5eq:nonlinear.tildeJ}), 
the symmetric additive synaptic noise, the symmetric random deletion, and the nonlinear synapse 
can be discussed in terms of the symmetric multiplicative synaptic noise. 
Therefore, first of all, we deal with the multiplicative synaptic noise. 

\subsection{Multiplicative synaptic noise}
\label{sec.sparse_noise.result.mul}

\begin{figure}
\epsfig{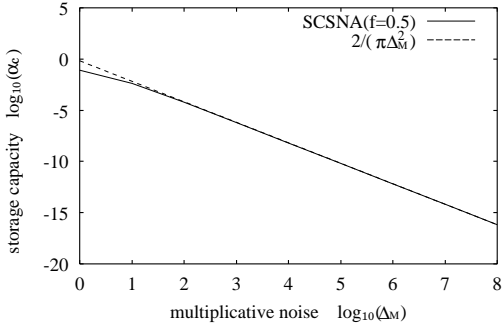}{8cm}
\caption[Dependence of storage capacity 
on multiplicative synaptic noise $\Delta_M^2$]
{
Dependence of storage capacity $\alpha_c$ 
on the multiplicative synaptic noise $\Delta_M^2$ 
at the firing rate $f=0.5$. 
Comparison of asymptote and the results from the SCSNA. 
}
\label{5fig:mul.ac-delta.f=0.5}
\end{figure}

Figure \ref{5fig:mul.ac-delta.f=0.5} shows 
the dependence of storage capacity on the multiplicative synaptic noise. 
As it is clear from Fig.\ref{5fig:mul.ac-delta.f=0.5}, 
storage capacity $\alpha_c$ is inversely proportional to 
the variance of the multiplicative synaptic noise $\Delta_M^2$, 
when the multiplicative synaptic noise is extremely large. 
Storage capacity $\alpha_c$ asymptotically approaches 
\begin{equation}
\alpha_c 
= \frac 2{\pi \Delta_M^2}, 
\label{5eq:mul.asymptote}
\end{equation}
(see Appendix \ref{appendix.critical_noise}). 
In the sparse limit where the firing rate is extremely small, 
it is known that storage capacity 
becomes $\alpha_c \simeq 1/(f|\log f|)$ 
\cite{Tsodyks1988,Amari1989,Buhmann1989,Perez-Vincente1989,Okada1996}. 

\begin{figure}
\epsfig{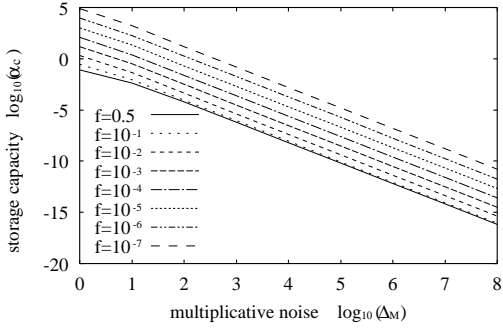}{8cm}
\caption[Dependence of storage capacity $\alpha_c$ 
on multiplicative synaptic noise $\Delta_M^2$]
{
Dependence of storage capacity $\alpha_c$ 
on the multiplicative synaptic noise $\Delta_M^2$. 
It can be confirmed that the order of the asymptote 
does not depend on the firing rate. 
}
\label{5fig:mul.ac-delta.all-f}
\end{figure}

Figure \ref{5fig:mul.ac-delta.f=0.5} shows 
the results from the SCSNA and the asymptote at the firing rate $f=0.5$. 
Figure \ref{5fig:mul.ac-delta.all-f} shows 
the results from the SCSNA at various firing rates. 
It can be confirmed 
that the order of the asymptote $O(\frac 1{\Delta_M^2})$ 
does not depend on the firing rate
from Fig.\ref{5fig:mul.ac-delta.all-f}. 

\subsection{Random deletion}
\label{sec.sparse_noise.result.random}

Next, we discuss the asymptote of the random deletion. 
The random deletion with the connecting rate $c$ can be transformed 
into the multiplicative synaptic noise by Eq.(\ref{5eq:cut.to.mul}). 
Hence, 
at the limit where the connecting rate $c$ is extremely small, 
storage capacity becomes 
\begin{equation}
\alpha_c
= \frac{2}{\pi \Delta_M^2}
= \frac {2c}{\pi (1-c)}
\to \frac 2{\pi} c, 
\label{5eq:cut.asymptote}
\end{equation}
according to the asymptote of the multiplicative synaptic noise 
in Eq.(\ref{5eq:mul.asymptote}). 
In the random deletion, 
the synapse efficiency $S_{eff}$, 
which is storage capacity per the connecting rate 
\cite{Okada1998,Sompolinsky1986b}, 
i.e., storage capacity per the input of one neuron, 
and defined as 
\begin{equation}
S_{eff} \equiv \frac{\alpha_c}{c},
\label{5eq:seff.definition}
\end{equation}
approaches a constant value as
\begin{equation}
S_{eff}=\frac{\alpha_c}{c}=\frac 2{\pi},
\end{equation}
according to Eq.(\ref{5eq:cut.asymptote}) 
at the limit where the connecting rate $c$ is extremely small. 

\begin{figure}
\epsfig{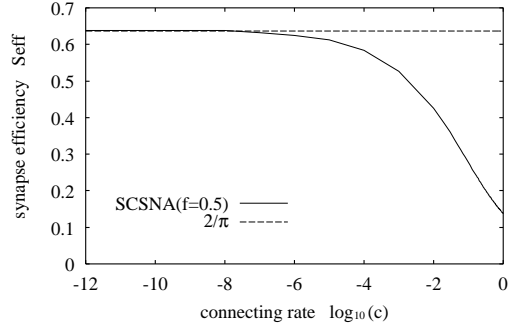}{8cm}
\caption[Dependence of the synapse efficiency $S_{eff}$ 
on random deletion with the connecting rate $c$]
{
Dependence of the synapse efficiency $S_{eff}$ 
on the random deletion with the connecting rate $c$ 
at the firing rate $f=0.5$. 
Comparison of asymptote and the results from the SCSNA. 
}
\label{5fig:cut.ac-delta.f=0.5}
\end{figure}

\begin{figure}
\epsfig{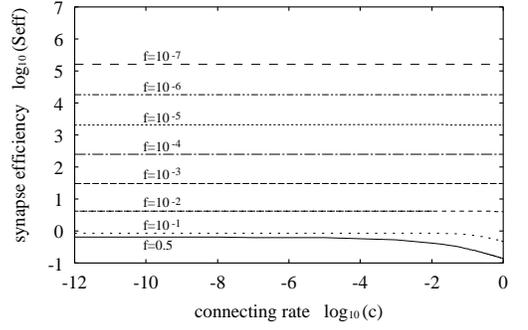}{8cm}
\caption[Dependence of the synapse efficiency $S_{eff}$ 
on random deletion with the connecting rate $c$ at various firing rates. ]
{
Dependence of the synapse efficiency $S_{eff}$ 
on the random deletion with the connecting rate $c$ at various firing rates. 
It can be confirmed that the order of the asymptote 
does not depend on the firing rate. 
}
\label{5fig:cut.ac-delta.all-f}
\end{figure}

Figure \ref{5fig:cut.ac-delta.f=0.5} shows 
the result from the SCSNA and the asymptote at the firing rate $f=0.5$. 
Figure \ref{5fig:cut.ac-delta.all-f} shows 
the results from the SCSNA at various firing rates. 
It can be confirmed that the order 
of the asymptote, $O(1)$ with respect to $c$, 
does not depend on the firing rate 
from Fig.\ref{5fig:cut.ac-delta.all-f}. 

\subsection{Systematic Deletion}
\label{sec.sparse_noise.result.systematic}


\subsubsection{Clipped synapse}
\label{sec.sparse_noise.result.systematic.clip}

Synapses within the range $-t<z<t$ are pruned 
by the nonlinear function of Eq.(\ref{5eq:nonlinear.clip}). 

\begin{figure}
\epsfig{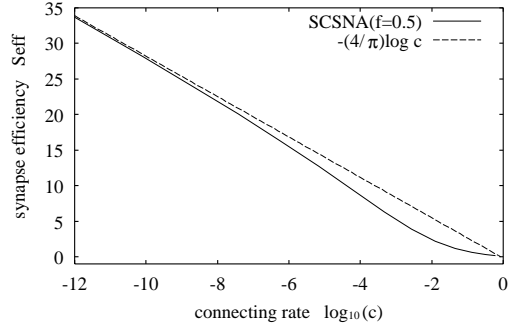}{8cm}
\caption[]
{
Dependence of the synapse efficiency $S_{eff}$ with the clipped synapse 
on the connecting rate $c$ at $f=0.5$. 
Comparison of the results from the SCSNA and asymptote. 
}
\label{5fig:clip.ac-delta.f=0.5}
\end{figure}

\begin{figure}
\epsfig{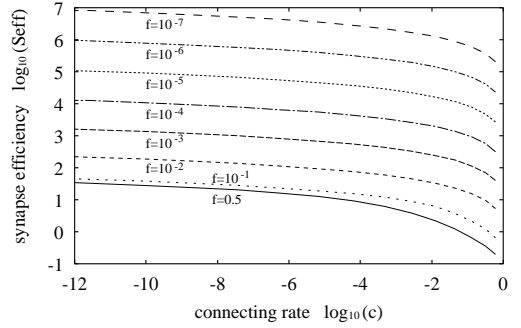}{8cm}
\caption[]
{
Dependence of the synapse efficiency $S_{eff}$ 
with the clipped synapse on the firing rate $f$. 
It can be confirmed that the order of the asymptote 
does not depend on the firing rate. 
}
\label{5fig:clip.ac-delta.all-f}
\end{figure}

The connection rate $c$ 
of the synaptic deletion in Eq.(\ref{5eq:nonlinear.T}) is given by, 
\begin{eqnarray}
c&=&\int_{ \{ z|f_1(z,t) \ne 0 \} } Dz
 =1-\erf (\frac{t}{\sqrt 2}) \nonumber \\
& \to & 
\sqrt{\frac {2}{\pi}} t^{-1} \exp (-\frac {t^2}2), \; \; t \to \infty, 
\label{5eq:clip.connect}
\end{eqnarray}
since the synaptic connection $T_{ij}$ before acting 
the nonlinear function of Eq.(\ref{5eq:nonlinear.T}) 
obeys the Gaussian distribution $N(0,1)$. 
Next, $J,\tilde{J}$ 
of Eqs.(\ref{5eq:nonlinear.J}) and (\ref{5eq:nonlinear.tildeJ}) become 
\begin{eqnarray}
J &=& 
      2 \int_t^\infty Dz \; z \sgn(z)
      = \sqrt{\frac 2{\pi}} \exp(-\frac {t^2}2) \nonumber \\
      & \to & t c, \; t \to \infty ,
      \label{5eq:clip.J} \\
{\tilde J}^2 &=& 
      2 \int_t^\infty Dz
      =1-\erf (\frac{t}{\sqrt 2})
      =c. 
      \label{5eq:clip.tildeJ}
\end{eqnarray}
Hence, 
the equivalent multiplicative synaptic noise $\Delta_M^2$ is obtained as, 
\begin{equation}
\Delta_M^2
   = \frac{{\tilde J}^2}{J^2}-1
   \to \frac 1{t^2c}, \; \; t \to \infty .
\label{5eq:clip.deltaM}
\end{equation}
The relationship of the pruning range $t$ and the connecting rate $c$ 
\begin{equation}
t^2 = -2 \log c,
\label{5eq:t.to.c}
\end{equation}
is obtained by taking the logarithm of Eq.(\ref{5eq:clip.connect})
at $t \to \infty$ limit. 
Therefore, at the limit where the equivalent connecting rate $c$ is extremely small, 
storage capacity $\alpha_c$ can be obtained 
\begin{equation}
\alpha_c = - \frac 4{\pi} c \log c,
\label{5eq:clip.asymptote}
\end{equation}
through Eqs.(\ref{5eq:mul.asymptote}), (\ref{5eq:clip.deltaM}), 
and (\ref{5eq:t.to.c}). 
The synapse efficiency becomes 
\begin{equation}
S_{eff}=\frac{\alpha_c}{c} = - \frac 4{\pi} \log c. 
\end{equation}
Figure \ref{5fig:clip.ac-delta.f=0.5} shows 
the results from the SCSNA and the asymptote at the firing rate $f=0.5$. 
Figure \ref{5fig:clip.ac-delta.all-f} shows 
the results from the SCSNA at various firing rates. 
It can be confirmed that the order of the asymptote $O(\log c)$ 
does not depend on the firing rate 
from Fig.\ref{5fig:clip.ac-delta.all-f}. 

\subsubsection{Minimal value deletion}
\label{sec.sparse_noise.result.systematic.minimal}

\begin{figure}
\epsfig{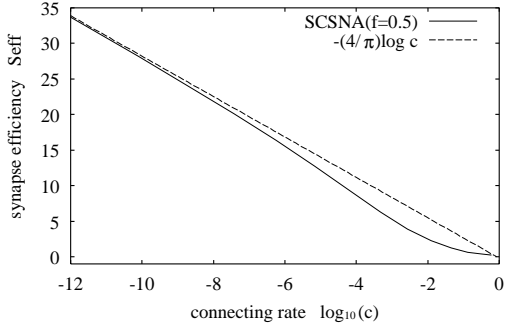}{8cm}
\caption[]
{
Dependence of the synapse efficiency $S_{eff}$ 
with the minimal value deletion on the connecting rate $c$ at time $f=0.5$. 
Comparison of the results from the SCSNA and the asymptote. 
}
\label{5fig:minimal.ac-delta.f=0.5}
\end{figure}

\begin{figure}
\epsfig{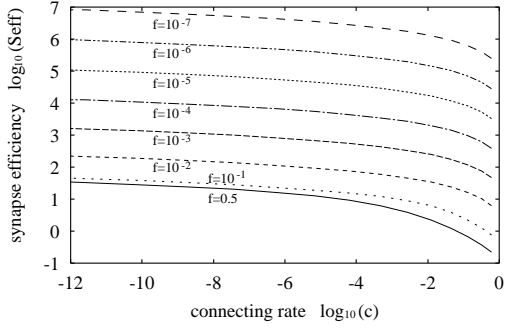}{8cm}
\caption[]
{
Dependence of the synapse efficiency $S_{eff}$ 
with the minimal value deletion on the firing rate $f$. 
It can be confirming that the order of the asymptote 
does not depend on the firing rate. 
}
\label{5fig:minimal.ac-delta.all-f}
\end{figure}

In a similar way, 
the equivalent multiplicative synaptic noise $\Delta_M^2$ 
of the systematic deletion of Eq.(\ref{5eq:nonlinear.minimal}) 
is obtained as follows, 
\begin{equation}
\Delta_M^2
   = \frac{{\tilde J}^2}{J^2}-1
   \to \frac 1{t^2c}, \; \; t \to \infty .
\label{5eq:minimal.deltaM}
\end{equation}
where
the connecting rate $c$ 
and $J,\tilde{J}$ 
of Eqs.(\ref{5eq:nonlinear.J}) and (\ref{5eq:nonlinear.tildeJ}) are 
\begin{eqnarray}
c
&=&
\int_{ \{ z|f_2(z,t) \ne 0 \} } Dz
=
1-\erf (\frac{t}{\sqrt 2}) \nonumber \\
& \to &
\sqrt{\frac {2}{\pi}} t^{-1} \exp (-\frac {t^2}2), \; \; t \to \infty, 
\label{5eq:minimal.connect} \\
J &=& 
      \sqrt{\frac 2{\pi}} t \exp(-\frac {t^2}2)
        +1-\erf (\frac{t}{\sqrt 2}) \nonumber \\
      & \to & \sqrt{\frac 2{\pi}} t \exp(-\frac {t^2}2), \; \; t \to \infty,
      \label{5eq:minimal.J} \\
{\tilde J}^2 &=& J, 
      \label{5eq:minimal.tildeJ}
\end{eqnarray}
respectively.
Hence, 
at the limit where 
the equivalent connecting rate $c$ is extremely small, 
storage capacity $\alpha_c$ 
and the synapse efficiency $S_{eff}$ can be obtained 
through Eqs.(\ref{5eq:mul.asymptote})，
(\ref{5eq:minimal.deltaM}), and (\ref{5eq:t.to.c}) 
as follows, 
\begin{eqnarray}
\alpha_c &=& -\frac 4{\pi} c \log c, \label{5eq:minimal.asymptote} \\
S_{eff} &=& - \frac 4 {\pi} \log c, 
\end{eqnarray}
respectively.
Figure \ref{5fig:minimal.ac-delta.f=0.5} 
shows the results from the SCSNA 
and the asymptote at the firing rate $f=0.5$. 
Figure \ref{5fig:minimal.ac-delta.all-f} shows 
the results from the SCSNA at various firing rates. 
It can be confirmed that the order of the asymptote 
does not depend on the firing rate 
from Fig.\ref{5fig:minimal.ac-delta.all-f}. 

\subsubsection{Compressed deletion}
\label{sec.sparse_noise.result.systematic.compress}

\begin{figure}
\epsfig{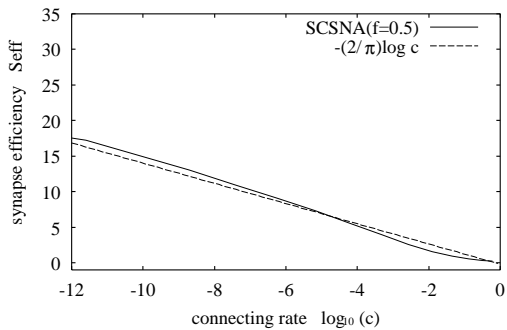}{8cm}
\caption[]
{
Dependence of the synapse efficiency $S_{eff}$ with the compressed deletion 
on the connecting rate $c$ at $f=0.5$. 
Comparison of the results from the SCSNA and the asymptote. 
}
\label{5fig:compress.ac-delta.f=0.5}
\end{figure}

\begin{figure}
\epsfig{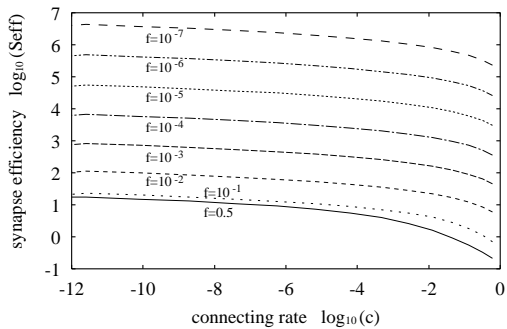}{8cm}
\caption[]
{
Dependence of the synapse efficiency $S_{eff}$ with the compressed deletion 
on the firing rate $f$. 
It can be confirming that the order of the asymptote 
does not depend on the firing rate. 
}
\label{5fig:compress.ac-delta.all-f}
\end{figure}

Again, in the similar way, 

The equivalent multiplicative synaptic noise $\Delta_M^2$ 
of the systematic deletion of Eq.(\ref{5eq:nonlinear.compress}) 
is given by 
\begin{equation}
\Delta_M^2
   = \frac{{\tilde J}^2}{J^2}-1
   \to \frac 2{t^2c}, \; \; t \to \infty, 
\label{5eq:compress.deltaM}
\end{equation}
where the connecting rate $c$ and 
$J,\tilde{J}$ of Eqs.(\ref{5eq:nonlinear.J}) 
and (\ref{5eq:nonlinear.tildeJ}) are 
\begin{eqnarray}
c&=&\int_{ \{ z|f_3(z,t) \ne 0 \} } Dz
 =1-\erf (\frac{t}{\sqrt 2}), \nonumber \\
& \to & 
\sqrt{\frac {2}{\pi}} t^{-1} \exp (-\frac {t^2}2), \; \; t \to \infty, 
\label{5eq:compress.connect} \\
J
&=&
c \label{5eq:compress.J} \\
{\tilde J}^2
&=& 
     \int_{-\infty}^\infty Dz \; f_2(z,t)^2
+t^2 \int_{-\infty}^\infty Dz \; f_1(z,t)^2 \nonumber \\
& &
\qquad -2t  \int_{-\infty}^\infty Dz \; f_1(z,t)f_2(z,t) \nonumber \\
&=&
     \biggl[ \sqrt{\frac 2\pi}t\exp (-\frac {t^2}2) 
             + 1-\erf (\frac t{\sqrt{2}}) \biggr] \nonumber \\
& &
+t^2 \biggl[ 1-\erf (\frac t{\sqrt{2}}) \biggr] 
-2 t \biggl[ \sqrt{\frac 2\pi} \exp (-\frac {t^2}2) \biggr] \nonumber \\
&\to&
\frac{2c}{t^2}, \; \; t \to \infty, \label{5eq:compress.tildeJ}
\end{eqnarray}
respectively. 
Here, we use a asymptotic expansion equation of the error function 
\begin{eqnarray}
\erf (x) = 1 - \frac{e^{-x^2}}{\sqrt{\pi}}
\left(\frac 1x- \frac 1{2x^3} + \frac 3{4x^5} - O(x^{-7}) \right), 
\end{eqnarray}
for $x \gg 1$. 
In order for the first term and the second term 
of Eq.(\ref{5eq:compress.tildeJ}) to be same order, 
the asymptotic expansion equation has taken 
the approximation of $O(t^{-3})$ and $O(t^{-5})$ respectively. 
The equivalent multiplicative synaptic noise 
in the case of systematic deletion becomes double 
that of the clipped synapse of Eq.(\ref{5eq:clip.deltaM}) 
and the minimal value deletion of Eq.(\ref{5eq:compress.deltaM}). 
Therefore, 
at the limit where the equivalent connecting rate $c$ is extremely small,
storage capacity $\alpha_c$ and the synapse efficiency $S_{eff}$ 
can be obtained 
through Eqs.(\ref{5eq:mul.asymptote})，
(\ref{5eq:compress.deltaM}), and (\ref{5eq:t.to.c}) 
as follows, 
\begin{eqnarray}
\alpha_c &=& - \frac 2{\pi} c \log c,
\label{5eq:compress.asymptote} \\
S_{eff} &=& - \frac 2{\pi} \log c, 
\end{eqnarray}
respectively. 
Figure \ref{5fig:compress.ac-delta.f=0.5} shows 
the results from the SCSNA and the asymptote at the firing rate $f=0.5$. 
Figure \ref{5fig:compress.ac-delta.all-f} shows 
the results by the SCSNA at various firing rates. 
It can be confirmed that the order of the asymptote $O(\log c)$
does not depend on the firing rate 
from Fig.\ref{5fig:compress.ac-delta.all-f}. 

\begin{figure}
\epsfig{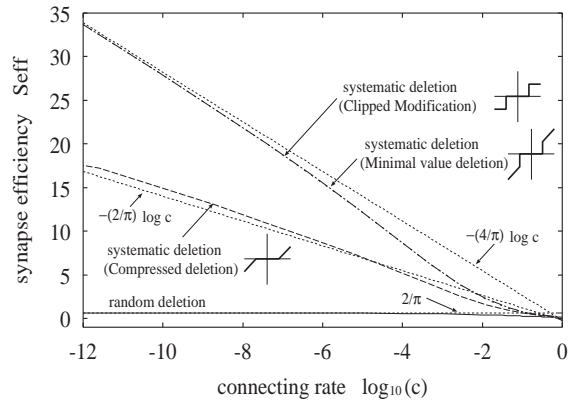}{8cm}
\caption[]
{
Comparison of the synapse efficiency with the random deletion 
and that with the systematic deletion at the firing rate $f=0.5$. 
}
\label{5fig:seff.conclusion}
\end{figure}

Figure \ref{5fig:seff.conclusion} shows 
the dependence of the synapse efficiency $S_{eff}$ 
on the connecting rate $c$ obtained by means of the SCSNA. 
Table \ref{5table:asymtotics} shows 
the asymptote of storage capacity 
with the random deletion and the systematic deletion. 
Hence, 
when using minimal value deletion as the simplest from of systematic deletion 
we found that the synapse efficiency 
in the case of systematic deletion becomes 
\begin{equation}
\frac{S_{eff}({\rm systematic \; deletion})}
{S_{eff}({\rm random \; deletion})}
=\frac{\frac 4{\pi}| \log c|}{\frac 2{\pi}}=2|\log c| , 
\end{equation}
thus 
we have shown analytically that 
the synapse efficiency in the case of systematic deletion 
diverges as $O(|\log c|)$ 
at the limit where the connecting rate $c$ is extremely small, 
and have shown that the synapse efficiency 
in the case of the systematic deletion becomes $2 |\log c|$ times 
as large as that of the random deletion. 

\begin{table}
\caption{
Asymptote of storage capacity 
by the random deletion and by the systematic deletion 
at the firing rate $f=0.5$. 
}
\label{5table:asymtotics}
\begin{center}
\begin{tabular}{ll|c}
\hline
Types of deletion & & Storage capacity \\
     & & (Asymptote) \\
\hline
random deletion & & $(2/\pi) c$ \\
systematic deletion & clipped synapse & $(4/\pi) c| \log c| $\\
& minimal value deletion & $(4/\pi) c| \log c|$ \\
& compressed deletion & $(2/\pi) c| \log c|$ \\
\hline
\end{tabular}
\end{center}
\end{table}

\begin{table}
\caption{
Asymptote of the synapse efficiency 
by the random deletion and the systematic deletion 
at the firing rate $f=0.5$. 
}
\label{5table:asymtotics.seff}
\begin{center}
\begin{tabular}{ll|c} 
\hline
Types of deletion & & Synapse efficiency \\
     & & (Asymptote) \\
\hline
random deletion & & $(2/\pi)$ \\
systematic deletion & clipped synapse & $(4/\pi) | \log c|$ \\
& minimal value deletion & $(4/\pi) | \log c|$ \\
& compressed deletion & $(2/\pi) | \log c|$ \\
\hline
\end{tabular}
\end{center}
\end{table}

\section{The memory performance under limited metabolic energy resources}

Until the previous section, 
we have discussed the effect of synaptic pruning 
by evaluating the synapse efficiency 
which is the memory capacity normalized by connecting rate $c$.
When the connecting rate is $c$, 
the synapse number per one neuron decreases to $cN$. 
Therefore, 
the synapse efficiency means 
the capacity per the input of one neuron. 
In the discussion by the synapse efficiency, 
the synapse number decreases when the connecting rate is small. 
\par
In addition to such the fixed neuron number criterion, 
a fixed synapse number criterion could be discussed. 
At the synaptic growth stage, 
it is natural to assume 
that metabolic energy resources are restricted. 
When metabolic energy resources are limited, 
it is also important 
that the effect of synaptic pruning is discussed 
under limited synapse number. 
Chechik et al. discussed 
the memorized pattern number per one synapse 
under a fixed synapse number criterion \cite{Chechik1998}．
They pointed out 
the existence of an optimal connecting rate 
under the fixed synapse number criterion 
and suggested an explanation of synaptic pruning as follows: 
synaptic pruning following over-growth can improve 
performance of a network with limited synaptic resources. 
\par
\begin{figure}[h]
\begin{center}
\includegraphics[width=0.8 \linewidth,keepaspectratio]{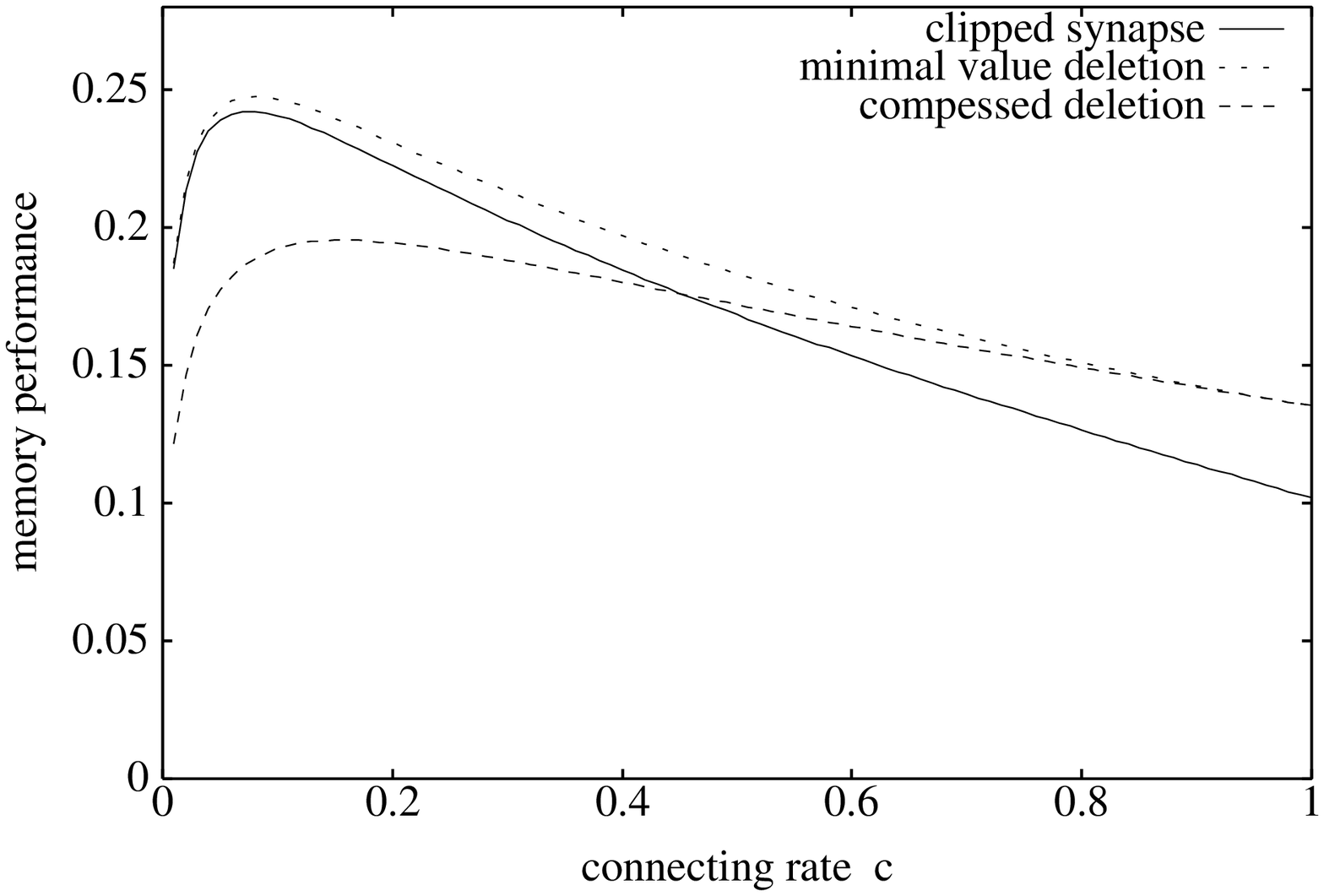}
\caption{Memory performance $\alpha_c/\sqrt{c}$ 
of networks with different number of neurons 
but the same total number of synapses 
as a function of the connecting rate $c$ 
in the case of clipped synapse (the solid line),
minimal value deletion (the dotted line),
and compressed deletion (the dashed line).}
\label{fig:metaboric}
\end{center}
\end{figure}

The synapse number is $N^2$ in the full connected case $c=1$. 
We consider the larger network with $M(>N)$ neurons. 
The synapse number in the lager networks 
with the connecting rate $c$ becomes $cM^2$. 
We can introduce the fixed synapse number criterion 
by considering a larger network 
which has $M$ neurons, i.e., 
\begin{equation}
N^2=cM^2,
\end{equation}
synapses at the connecting rate $c$.
The memorized pattern number per one synapse becomes 
\begin{equation}
\frac{p_c}{cM^2}
= \frac {p_c}{N^2}
= \frac{\alpha_c}N
= \frac{\alpha_c}{\sqrt{c}M},
\end{equation}
where 
the critical memorized pattern number $p_c$ is 
\begin{equation}
p_c=\alpha_c N.
\end{equation}
We define the coefficient $\alpha_c/\sqrt{c}$ as memory performance. 
We discuss the effect of synaptic pruning by the memory performance. 
Under limited metabolic energy resources, 
the optimal strategy is the maximization of the memory performance. 
Chechik et al. showed 
the existence of the optimal connecting rate 
which maximize memory performance \cite{Chechik1998}. 
The memory performance can be calculated by 
normalising
the capacity, which is given by solving the order-parameter equations, 
with $\sqrt{c}$. 
Figure \ref{fig:metaboric} shows 
the dependence of the memory performance on the connecting rate 
in three types of pruning. 
It is confirmed 
that there are the optimal values 
which maximized the memory performance by each deletions. 
The optimal connecting rates 
of clipped synapse, minimal value deletion, and compressed deletion 
are $c=0.036,0.038,0.084$, respectively. 
This interesting fact may imply that 
the memory performance is improved without heavy pruning. 
These optimal values agree with the computer simulation results 
given by Chechik et al \cite{Chechik1998}.

\section{Conclusion}
\label{sec.sparse_noise.conclusion}

We have analytically discussed the synapse efficiency, 
which we regarded as the auto-correlation-type associative memory, 
to evaluate the effect of the pruning following over-growth. 
Although Chechik et al. pointed out 
that the synapse efficiency is increased by the systematic deletion, 
this is qualitatively obvious 
and the increase in the synapse efficiency 
should also be discussed quantitatively. 
At the limit where the multiplicative synaptic noise is extremely large, 
storage capacity $\alpha_c$ is inversely 
as the variance of the multiplicative synaptic noise $\Delta_M^2$. 
From this result, 
we analytically obtained that 
the synapse efficiency in the case of the systematic deletion 
diverges as $O(|\log c|)$
at the limit where the connecting rate $c$ is extremely small. 

On the other hand, it is natural to assume 
that metabolic energy resources are restricted 
at the synaptic growth stage. 
When metabolic energy resources are limited, 
i.e., synapse number is limited, 
the optimal connecting rate, which maximize memory performance, exists.
These optimal values are in agreement 
with the results given by Chechik et al \cite{Chechik1998}.

In the correlation learning, which can be considered a prototype 
of any other learning rules, 
various properties can be analyzed quantitatively. 
The asymptote of synapse efficiency in the model with another learning rule 
can be discussed in a similar way. 
As our future work, we plan to further discuss 
these properties while taking into account various considerations 
regarding related physiological knowledge. 
\\

\begin{center}
{\small \bf ACKNOWLEDGEMENTS}
\end{center}

We thank the referee for the helpful comments, 
especially for the comment concerning the derivation of Appendix 
\ref{appendix.critical_noise}. 
This work was partially supported by a Grant-in-Aid 
for Scientific Research on Priority Areas No. 14084212, 
for Scientific Research (C) No. 14580438, 
for Encouragement of Young Scientists (B) No. 14780309 and 
for Encouragement of Young Scientists (B) No. 15700141 
from the Ministry of Education, Culture, Sports, Science 
and Technology of Japan.

\appendix

\section{SCSNA for additive synaptic noise}
\label{appendix.scsna}

Derivations of the order-parameter 
Eqs.(\ref{eq:order_parameter_equation_m})
-(\ref{eq:effevtive_response_function}) are given here. 
The SCSNA starts from the fixed-point equations 
for the dynamics of the $N$-neuron network 
shown as Eq.(\ref{5eq:sparse.equilibrium}). 
The random memory patterns are generated 
according to the probability distribution 
of Eq.(\ref{5eq:sparse.pattern}). 
The syanaptic connections are given by Eq.(\ref{5eq:add.noise.J}). 
The asymmetric additive synaptic noise, $\delta_{ij}$ and $\delta_{ji}$ 
are independently generated according to the probability distribution 
of Eq.(\ref{5eq:add.noise.pdf}). 
Moreover, we can analyze a more general case, 
where $\delta_{ij}$ and $\delta_{ji}$ have an arbitary correlation such that 
\begin{equation}
{\rm Cov}[\delta_{ij}, \delta_{ji}]
  = k_\delta \frac{J^2}{N(1-a^2)^2} \Delta_A^2, \quad
  -1 \leq k_\delta \leq 1.
\end{equation}
In this general case, 
the symmetric and the asymmetric additive synaptic noise 
correspond to $k_\delta =1$ and $k_\delta =0$, respectively. 
Here, we assume the probability distribution of the additive syanptic noise 
is normal distribution 
$\delta_{ij} \sim N(0, \frac{J^2}{N(1-a^2)^2} \Delta_A^2)$. 
However, any probability distributions, which have same average and variance, 
can be discussed by the central limit theorem 
in the similar way of the following disscussion. 
Defining the loading rate as $\alpha =p/N$, 
we can write the local field $h_i$ for neuron $i$ as 
\begin{eqnarray}
h_i 
&\equiv& \sum_{j\ne i}^N J_{ij} x_j \nonumber \\
&=& J \sum_{\mu=1}^{\alpha N} (\xi_i^\mu-a) m_\mu +
    \sum_{j \neq i}^N \delta_{ij} x_j - J \alpha x_i,
\label{ap:add.h_i}
\end{eqnarray}
where $m_\mu$ is the overlap between the stored pattern $\vec{\xi}^\mu$ 
and the equilibrium state $\vec{x}$ defined by 
\begin{equation}
m_\mu = \frac{1}{N(1-a^2)} \sum_{i=1}^N (\xi^\mu _i-a) x_i. 
\label{eq:appA_definition_overlap}
\end{equation}
The second term including $x_j=F(h_j+h)$ in Eq.(\ref{ap:add.h_i}) 
depends on $\delta_{ji}$. 
The $\delta_{ij}$ dependences of $x_j$ are extracted from $x_j$, 
\begin{equation}
x_j = x^{(\delta_{ji})}_j + \delta_{ji} x_i x'{}^{(\delta_{ji})}_j, 
    \label{ap:add.eq.x_j}
\end{equation}
where 
\begin{eqnarray}
x ^{(\delta_{ji})}_j &=& F (h_j - \delta_{ji} x_i), \\
x'{}^{(\delta_{ji})}_j &=& F'(h_j - \delta_{ji} x_i).
\end{eqnarray}
Substituting Eq.(\ref{ap:add.eq.x_j}) into Eq.(\ref{ap:add.h_i}), 
the local field $h_i$ can be expressed as 
\begin{eqnarray}
h_i 
  &=& J \sum_{\mu=1}^{\alpha N} (\xi_i^\mu-a) m_\mu + 
  \sum_{j \neq i}^N \delta_{ij} x^{(\delta_{ji})}_j -
  J \alpha x_i \nonumber \\
  & & + x_i \sum_{j \neq i}^N \delta_{ij} 
  \delta_{ji} x'{}^{(\delta_{ji})}_j. 
  \label{ap:add.eq.h_i}
\end{eqnarray}
We assume that 
Eq.(\ref{ap:add.eq.h_i}) and $x_i=F(h_i+h)$ can be solved 
by using the effective response function $\tilde{F}(u)$ as, 
\begin{equation}
x_i = \tilde{F} \biggl( J \sum_{\mu=1}^p (\xi_i^\mu-a) m_\mu 
  + \sum_{j \neq i}^N \delta_{ij} x^{(\delta_{ji})}_j \biggr) .
\label{ap:add.overlap.xi}
\end{equation}
Let $\vec{\xi}^1$ be the target pattern to be retrieved. 
Therefore, 
we can assume that $m_1=O(1)$ and $m_\mu =O(1/\sqrt{N}),\mu >1$. 
Then we can use the Taylor expansion to obtain 
\begin{eqnarray}
m_\mu 
&=& \frac{1}{N(1-a^2)} \sum_{i=1}^N (\xi_i^\mu-a) x_i^{(\mu)} 
  \nonumber \\
  & & +
  \frac{J}{N(1-a^2)} \sum_{i=1}^N (\xi^\mu_i - a)^2 m_\mu x'{}^{(\mu)}_i 
  \nonumber \\
&=& \frac{1}{N(1-a^2)} \sum_{i=1}^N
  (\xi_i^\mu-a) x_i^{(\mu)} + J U m_\mu 
  \nonumber \\
&=& \frac{1}{N(1-a^2)(1-J U)} 
\sum_{i=1}^N (\xi_i^\mu-a) x_i^{(\mu)}, 
\label{ap:add.mu} 
\end{eqnarray}
by substituting Eq.(\ref{ap:add.overlap.xi}) 
into the overlap defined by Eq.(\ref{eq:appA_definition_overlap}), 
where 
\begin{eqnarray}
x^{(\mu)}_i &=&
\tilde{F} \biggl( J \sum_{\nu \neq \mu}^{\alpha N} (\xi_i^\nu-a) m_\nu + 
\sum_{j \neq i}^N \delta_{ij} x^{(\delta_{ji})}_j \biggr), \\
x'{}^{(\mu)}_i &=&
\tilde{F}' \biggl( J \sum_{\nu \neq \mu}^{\alpha N} (\xi_i^\nu-a) m_\nu + 
\sum_{j \neq i}^N \delta_{ij} x^{(\delta_{ji})}_j \biggr), \\
U &=& \frac{1}{N} \sum_{i=1}^N x'{}^{(\mu)}_i.
\label{eq:appA_additive_U}
\end{eqnarray}
Equations (\ref{ap:add.eq.h_i}) and (\ref{ap:add.mu}) give 
the following expression for the local field: 
\begin{eqnarray}
h_i 
&=& J (\xi^1_i-a) m_1 \nonumber \\
  & & + 
  \alpha \left[\frac{J^2}{1-JU} 
  +k_\delta \frac{J^2}{(1-a^2)^2} \Delta_A^2 \right] U x_i \nonumber \\
  & & + \frac{J}{N(1-a^2)(1-JU)} \sum_{j \neq i}^N \sum_{\mu = 2}^{\alpha N}
  (\xi^\mu_i-a) (\xi^\mu_j-a) x_j^{(\mu)} \nonumber \\
  & & + 
  \sum_{j \neq i}^N \delta_{ij} x^{(\delta_{ji})}_j.  
  \label{ap:add.eq.h_i_final}
\end{eqnarray}
Note that the second term in Eq.(\ref{ap:add.eq.h_i_final}) 
denotes the effective self-coupling term. 
The third and the last terms are 
summations of uncorrelated random variables, with mean 0 and variance, 
\begin{eqnarray}
& &\frac{J^2}{N^2(1-a^2)^2(1-JU)^2} \nonumber \\
& &\quad \times \sum_{j \neq i}^N \sum_{\mu = 2}^{\alpha N} 
(\xi_i^\mu-a)^2 (\xi_j^\mu-a)^2 (x_j^{(\mu)})^2 \nonumber \\
& & \qquad \qquad = \frac{\alpha J^2}{(1-JU)^2} q, 
\label{ap:add.eq.sigma1}
\end{eqnarray}
\begin{equation}
\sum_{j \neq i}^N \delta_{ij}^2 (x_j^{(\delta_{ji})})^2
= \frac {J^2}{(1-a^2)^2}\Delta_A^2 q, 
\label{ap:add.eq.sigma2}
\end{equation}
respectively.
The cross term of these terms have vanished. 
Thus, we finally obtain 
\begin{eqnarray}
h_i &=& J (\xi^1_i-a) m_1 + \sigma z_i \nonumber \\
& &+ \biggl[ \frac{\alpha J^2}{1-JU} +
    k_\delta \frac {J^2}{(1-a^2)^2} \Delta_A^2 \biggr] U x_i  \\
\sigma^2 &=& \frac{\alpha J^2 q}{(1-JU)^2} + 
       \frac {J^2}{(1-a^2)^2}\Delta_A^2 q,
\label{ap:sigma.additive_noise}
\end{eqnarray}
from Eqs.(\ref{ap:add.eq.sigma1}) and (\ref{ap:add.eq.sigma2}), 
where $z_i \sim N(0,1)$.
Equation (\ref{eq:order_parameter_equation_sigma}) is 
given by Eq.(\ref{ap:sigma.additive_noise}). 
Finally, 
after rewriting $\xi^1_i \to \xi$, $m_1 \to m$, $z_i \to z$, 
and $x_i \to Y(z;\xi )$, 
the results of the SCSNA for the additive synaptic noise 
are summarized by the order-parameter equations 
of Eqs.(\ref{eq:order_parameter_equation_m})
-(\ref{eq:order_parameter_equation_U}) as, 
\begin{eqnarray}
m &=& \frac{1}{1-a^2} \int Dz < (\xi -a) Y(z;\xi ) >_\xi ,\nonumber \\
q &=& \int Dz < Y(z;\xi )^2 >_\xi ,\nonumber \\
U &=& \frac{1}{\sigma} \int Dz \; z < Y(z;\xi ) >_\xi , \nonumber 
\end{eqnarray}
where the effective response function $Y(z;\xi )$ becomes 
\begin{eqnarray}
Y(z;\xi ) 
&=& F \biggl( J (\xi-a) m + \sigma z + h \nonumber \\
& & + \biggl[ \frac{\alpha J^2}{1-JU} +
    k_\delta \frac{J^2}{(1-a^2)^2} \Delta_A^2 \biggr] U Y(z;\xi ) \biggr). 
    \nonumber \\
& &
\label{ap:effective_response_function}
\end{eqnarray}
The effective response function of Eq.(\ref{eq:effevtive_response_function}) 
can be obtained by substituting $k_\delta =1$ 
into Eq.(\ref{ap:effective_response_function}).

\section{Equivalence among three types of noise}
\label{appendix.equivalent_noise}

The multiplicative synaptic noise, 
the random synaptic deletion, and the nonlinear synapse 
can be discussed in the similar manner to Appendix \ref{appendix.scsna}.

\subsection{Multiplicative synaptic noise}
\label{appendix.multiplicative_noise}

Derivations of the equivalent noise 
Eq.(\ref{5eq:add.to.mul}) is given here. 
We can also analyze 
by a similar manner to the analysis of the additive synaptic noise. 
The syanaptic connections are given by Eq.(\ref{5eq:mul.noise.J}). 
The asymmetric multiplicative synaptic noise, 
$\varepsilon_{ij}$ and $\varepsilon_{ji}$ are independently generated 
according to the probability distribution of Eq.(\ref{5eq:mul.noise.pdf}). 
We analyze a more general case, 
where $\delta_{ij}$ and $\delta_{ji}$ have an arbitary correlation such that 
\begin{equation}
\mbox{Cov}[\varepsilon_{ij}, \varepsilon_{ji}]
  = k_\varepsilon \Delta_M^2, \quad
  -1 \leq k_\varepsilon \leq 1.
\end{equation}
In this general case, the symmetric and the asymmetric multiplicative synaptic noise 
correspond to $k_\varepsilon =1$ and $k_\varepsilon =0$, respectively. 
Here, we assume the probability distribution of the multiplicative synaptic noise 
is normal distribution 
$\varepsilon_{ij} \sim N(0, \Delta_M^2)$. 
The local field $h_i$ for neuron $i$ becomes 
\begin{eqnarray}
h_i &=& \sum_{\mu=1}^{\alpha N} (\xi_i^\mu-a) m_\mu \nonumber \\
  & & + \frac{1}{N(1-a^2)} \sum_{\mu = 1}^{\alpha N} \sum_{j \neq i}^N 
     \varepsilon_{ij} (\xi_i^\mu-a) (\xi_j^\mu-a) x_j \nonumber \\
  & & - \alpha x_i,
\label{ap:mul.h_i}
\end{eqnarray}
where $m_\mu$ is the overlap 
defined by Eq.(\ref{eq:appA_definition_overlap}). 
The second term including $x_j=F(h_j+h)$ in Eq.(\ref{ap:mul.h_i}) 
depends on $\varepsilon_{ji}$. 
The $\varepsilon_{ij}$ dependences and $\xi_j^\mu$ dependences of $x_j$ 
are extracted from $x_j$, 
\begin{eqnarray}
x_j &=& x^{(\mu)(\varepsilon_{ji})}_j
  +  h_j^{\{ \mu,\varepsilon_{ji} \}} x'{}^{(\mu)(\varepsilon_{ji})}_j, 
    \label{ap:mul.eq.x_j} \\
h_j^{\{ \mu,\varepsilon_{ji} \}} 
  &=& (\xi^\mu_j-a) m_\mu \nonumber \\
  & & + \frac{1}{N(1-a^2)}
  \sum_{k \neq j}^N  \varepsilon_{jk} (\xi_j^\mu-a) (\xi_k^\mu-a) x_k 
  \nonumber \\
  & &+\frac{\varepsilon_{ji}}{N(1-a^2)} \sum_{\nu \neq \mu}^{\alpha N}
  (\xi_j^\nu-a) (\xi_i^\nu-a) x_i \nonumber \\
  & &+\frac{\varepsilon_{ji}}{N(1-a^2)} (\xi_j^\mu-a) (\xi_i^\mu-a) x_i,
  \label{ap:mul.eq.h_j}
\end{eqnarray}
where 
\begin{eqnarray}
x ^{(\mu)(\varepsilon_{ji})}_j&=&F (h_j - h_j^{\{ \mu,\varepsilon_{ji} \}}), \\
x'{}^{(\mu)(\varepsilon_{ji})}_j&=&F'(h_j - h_j^{\{ \mu,\varepsilon_{ji} \}}).
\end{eqnarray}
We assume that 
Eq.(\ref{ap:mul.h_i}) and $x_i=F(h_i+h)$ can be solved 
by using the effective response function $\tilde{F}(u)$ as, 
\begin{eqnarray}
& &x_i = \tilde{F}(\sum_{\mu=1}^p (\xi_i^\mu-a) m_\mu \nonumber \\
& &+ \frac{1}{N(1-a^2)} \sum_{\mu = 1}^{\alpha N} \sum_{j \neq i}^N 
  \varepsilon_{ij} (\xi_i^\mu-a) (\xi_j^\mu-a) x^{(\mu)(\varepsilon_{ji})}_j).
  \nonumber \\
& & 
  \label{ap:mul.overlap.xi}
\end{eqnarray}
Let $\vec{\xi}^1$ be the target pattern. 
We substitute Eq.(\ref{ap:mul.overlap.xi}) 
into the overlap defined by Eq.(\ref{eq:appA_definition_overlap}) 
and expand the resultant expression by $(\xi_i^\mu -a)m_\mu$ $(\mu >1)$,
which has the order of $O(1/\sqrt{N})$. 
This leads to 
\begin{equation}
m_\mu 
= \frac{1}{N(1-a^2)(1-U)} \sum_{i=1}^N (\xi_i^\mu-a) x_i^{(\mu)}, 
\label{ap:mul.mu} 
\end{equation}
where 
\begin{eqnarray}
& &x^{(\mu)}_i =
  \tilde{F} \biggl( \sum_{\nu \neq \mu}^{\alpha N} (\xi_i^\nu-a) m_\nu 
  \nonumber \\
& & + \frac{1}{N(1-a^2)} \sum_{\nu \neq \mu}^{\alpha N} \sum_{j \neq i}^N 
  \varepsilon_{ij} (\xi_i^\nu-a) (\xi_j^\nu-a) x^{(\nu)(\varepsilon_{ji})}_j
  \biggr), \nonumber \\
& & 
\end{eqnarray}
and $U$ is defined by the similar way of Eq.(\ref{eq:appA_additive_U}) 
in the case of the additive synaptic noise. 
Equations (\ref{ap:mul.h_i}),(\ref{ap:mul.eq.x_j}) and (\ref{ap:mul.mu}) give 
\begin{eqnarray}
h_i
&=& (\xi^1_i-a) m_1 + 
  \alpha \left[\frac{1}{1-U} +
  k_\varepsilon \Delta_M^2 \right] U x_i
  \nonumber \\
  & & + \frac{1}{N(1-a^2)(1-U)} \sum_{j \neq i}^N \sum_{\mu = 2}^{\alpha N}
  (\xi^\mu_i-a) (\xi^\mu_j-a) x_j^{(\mu)} \nonumber \\
  & & + \frac{1}{N(1-a^2)} \sum_{\mu = 1}^{\alpha N} \sum_{j \neq i}^N 
  \varepsilon_{ij} (\xi_i^\mu-a) (\xi_j^\mu-a) x^{(\mu)(\varepsilon_{ji})}_j.
  \nonumber \\
  \label{ap:mul.eq.h_i_final}
\end{eqnarray}
The third and last terms can be regarded as the noise terms. 
The variance of the noise terms becomes 
\begin{equation}
\sigma^2=
\frac{\alpha q}{(1-U)^2} + \alpha \Delta_M^2 q. 
\label{ap:sigma.multiplicative_noise}
\end{equation}
Thus, after rewriting $\xi_i^1 \to \xi$ and $m_1 \to m$, 
we obtain the effective response function: 
\begin{eqnarray}
Y(z;\xi ) &=&
F \biggl( (\xi -a) m + \sigma z +h \nonumber \\
& & + \alpha \left[\frac{1}{1-U} + k_\varepsilon \Delta_M^2 \right] U Y(z;\xi ) 
\biggr). \nonumber \\
\label{ap:mul.Y.final}
& &
\end{eqnarray}
Finally, 
the equivalence between the multiplicative synaptic noise 
and the additive synaptic noise is obtained as follows, 
\begin{eqnarray}
J &=& 1, \\
\Delta_A^2 &=& \alpha (1-a^2)^2 \Delta_M^2, \label{ap:mul2add} \\
k_\delta &=& k_\varepsilon, 
\end{eqnarray}
by comparing 
Eqs.(\ref{ap:sigma.multiplicative_noise}) and (\ref{ap:mul.Y.final}) 
to Eqs.(\ref{ap:sigma.additive_noise}) 
and (\ref{ap:effective_response_function}).

\subsection{Random deletion}
\label{appendix.random_deletion}

Derivations of the equivalent noise 
Eq.(\ref{5eq:cut.to.mul}) is given here. 
The random deletion has similar effects 
to the multiplicative synaptic noise. 
Therefore, we analyse by a similar way to 
the analysis of the multiplicative synaptic noise. 
The syanaptic connections are given by Eq.(\ref{5eq:random.cut.J}). 
The asymmetric cut coefficients are independently generated 
according to the probability distribution of Eq.(\ref{5eq:cut.pdf}). 
We analyze a more general case, 
where $c_{ij}$ and $c_{ji}$ have an arbitary correlation such that 
\begin{eqnarray}
\mbox{Cov}[c_{ij},c_{ji}] &=& k_c \mbox{Var}[c_{ij}], \quad
  -1 \leq k_c \leq 1 \\
\mbox{Var}[c_{ij}] &=& \mbox{E}[(c_{ij})^2] - (\mbox{E}[c_{ij}])^2 
  \nonumber \\
&=& c(1-c).
\end{eqnarray}
In this general case, the symmetric and asymmetric random deletion 
correspond to $k_c =1$ and $k_c =0$, respectively. 
According to a similar analysis of the multiplicative synaptic noise, 
the local field becomes 
\begin{eqnarray}
h_i 
&=& (\xi^1_i-a) m_1 + 
  \alpha \left[\frac{1}{1-U} + \frac{k_c(1-c)}{c} \right] U x_i 
  \nonumber \\
  & & + \frac{1}{N(1-a^2)(1-U)} \sum_{j \neq i}^N \sum_{\mu = 2}^{\alpha N}
  (\xi^\mu_i-a) (\xi^\mu_j-a) x_j^{(\mu)} \nonumber \\
  & & + \frac{1}{Nc(1-a^2)} \sum_{\mu = 1}^{\alpha N} \sum_{j \neq i}^N 
  (c_{ij}-c) \nonumber \\
  & & \qquad \qquad \times (\xi_i^\mu-a) (\xi_j^\mu-a) x^{(\mu)(c_{ji})}_j, 
  \label{ap:cut.eq.h_i_final}
\end{eqnarray}
where 
\begin{eqnarray}
& &x ^{(\mu)(c_{ji})}_j = F (h_j - h_j^{\{ \mu,c_{ji} \}}), \\
& & x^{(\mu)}_i =
  \tilde{F} \biggl( \sum_{\nu \neq \mu}^{\alpha N} (\xi_i^\nu-a) m_\nu + 
  \frac{1}{Nc(1-a^2)} \nonumber \\
  & & \quad \times \sum_{\nu \neq \mu}^{\alpha N} \sum_{j \neq i}^N 
  (c_{ij}-c) (\xi_i^\nu-a) (\xi_j^\nu-a) x^{(\nu)(c_{ji})}_j \biggr) , 
  \nonumber \\
& & \\
& &h_j^{\{ \mu,c_{ji} \}} = (\xi^\mu_j-a) m_\mu 
  + \frac{1}{Nc(1-a^2)} \nonumber \\
& &\quad  \times 
  \sum_{k \neq j}^N  (c_{jk}-c) (\xi_j^\mu-a) (\xi_k^\mu-a) x_k 
  \nonumber \\
& &\quad + \frac{c_{ji}-c}{Nc(1-a^2)} \sum_{\nu \neq \mu}^{\alpha N}
  (\xi_j^\nu-a) (\xi_i^\nu-a) x_i \nonumber \\
& &\quad + \frac{c_{ji}-c}{Nc(1-a^2)} (\xi_j^\mu-a) (\xi_i^\mu-a) x_i,
  \label{ap:cut.eq.h_j}
\end{eqnarray}
and $U$ is defined by Eq.(\ref{eq:appA_additive_U}) similarly. 
The variance of the noise term is given by 
\begin{equation}
\sigma^2=
\frac{\alpha q}{(1-U)^2} + \alpha \frac{1-c}{c} q. 
\label{ap:sigma.random_deletion}
\end{equation}
Thus, after rewriting $\xi_i^1 \to \xi$ and $m_1 \to m$, 
the effective response function becomes 
\begin{eqnarray}
Y(z;\xi ) &=&
F \biggl( (\xi -a) m + \sigma z +h \nonumber \\
& & + \alpha \left[\frac{1}{1-U} + \frac{k_c(1-c)}{c} \right] U Y(z;\xi ) 
\biggr). \nonumber \\
\label{ap:random_deletion.Y}
& &
\end{eqnarray}
Finally, 
the equivalence between random deletion 
and the additive synaptic noise is obtained as follows, 
\begin{eqnarray}
J &=& 1, \\
\Delta_A^2 &=& \alpha (1-a^2)^2 \frac{1-c}{c}, 
\label{ap:cut2add} \\
k_\delta &=& k_c, 
\end{eqnarray}
by comparing 
Eqs.(\ref{ap:sigma.random_deletion}) and (\ref{ap:random_deletion.Y}) 
to Eqs.(\ref{ap:sigma.additive_noise}) 
and (\ref{ap:effective_response_function}). 
Substituting Eq.(\ref{ap:cut2add}) into Eq.(\ref{ap:mul2add}), 
we obtain the equivalence of Eq.(\ref{5eq:cut.to.mul}).

\subsection{Nonlinear synapse}
\label{appendix.nonlinear_synapse}

Derivations of the equivalent noise 
Eq.(\ref{5eq:nonlinear.to.mul}) is given here. 
The effect of the nonlinear synapse can be separated into 
a signal part and a noise part. 
The noise part can be regarded as the additive synaptic noise. 

The systematic deletion of synaptic connections can be achieved 
by introducing synaptic noise with an appropriate nonlinear function $f(x)$ 
\cite{Sompolinsky1986b}. 
Note that $T_{ij}$ obeys the normal distribution $N(0,1)$ 
for $p=\alpha N \to \infty$. 
According to this naive S/N analysis \cite{Okada1998}, 
we can write the connections as 
\begin{eqnarray}
J_{ij} &=& \frac{\sqrt{p}}{N} f(T_{ij}) \nonumber \\
       &=& \frac{J}{N(1-a^2)} \sum_{\mu = 1}^{p}
           (\xi_{i}^{\mu}-a) (\xi_{j}^{\mu}-a) \nonumber \\
       & & - \biggl[ \frac{\sqrt{p}}{N} f(T_{ij}) 
           -\frac{J}{N(1-a^2)}
           \sum_{\mu = 1}^{p} (\xi_{i}^{\mu}-a) (\xi_{j}^{\mu}-a) \biggr] 
           \nonumber \\
       &=& \frac{\sqrt{p}}{N} \{J T_{ij} -[f(T_{ij})-J T_{ij}]\}. 
\label{ap:nonlinear.naive_sn.J}
\end{eqnarray}
The following derivation suggests that 
the residual overlap $m_\mu$ for the first term 
in Eq.(\ref{ap:nonlinear.naive_sn.J}) is enhanced 
by a factor of $1/(1-JU)$, while any enhancement to the last part is canceled 
because of the subtraction.
It also implies that the last part corresponds to the synaptic noise. 
For the SCSNA of the nonlinear synapse, 
we can analyze by a similar manner of the analysis of the additive synaptic noise. 
We obtain the local field: 
\begin{eqnarray}
h_i
&=& J (\xi_i^1-a) m_1 + \alpha \left[ \frac{J^2}{1-JU} 
+ (\tilde{J^2}-J^2)\right] U x_i \nonumber \\
& & + \frac{\sqrt{p}}{N} \sum_{j \neq i}
  \left[f(T_{ij}) - J T_{ij} \right] x^{(T_{ji})}_j \nonumber \\
& & + \frac{J}{N(1-a^2)(1-JU)}
    \sum_{\mu=2}^p \sum_{j \neq i}^N (\xi_i^\mu-a) (\xi_j^\mu-a) x_j^{(\mu)},
    \nonumber \\
& &
    \label{ap:nl.eq.h_i_final}
\end{eqnarray}
where 
\begin{eqnarray}
x^{(\mu)}_i 
&=& 
  \tilde{F} \biggl( J \sum_{\nu \neq \mu}^p (\xi_i^\nu-a) m_\nu 
  \nonumber \\
& & + \frac{\sqrt{p}}{N} \sum_{j \neq i}^N 
  [f(T^{(\mu)}_{ij}) - J T^{(\mu)}_{ij}] x^{(T_{ji})}_j \biggr), 
\label{ap:nonlinear.x_mu} \\
x^{(T_{ji})}_j 
  &=& F \biggl( h_j - \frac{\sqrt{p}}{N} [f(T_{ji})-J T_{ji}] x_i) \biggr) ,\\
T^{(\mu )}_{ij} &=& \frac{1}{\sqrt{p}(1-a^2)}
  \sum_{\nu \neq \mu}^p (\xi_i^\nu-a) (\xi_j^\nu-a), 
\end{eqnarray}
and $U$ is defined by Eq.(\ref{eq:appA_additive_U}) similarly. 
The variance of the noise term is given by 
\begin{equation}
\sigma^2=
  \frac{\alpha J^2 q}{(1-JU)^2}q 
+ \alpha (\tilde{J^2}-J^2). 
\label{ap:nonlinear.sigma}
\end{equation}
Thus, after rewriting $\xi_i^1 \to \xi$ and $m_1 \to m$, 
The effective response function becomes 
\begin{eqnarray}
Y(z;\xi ) &=&
F \biggl(
J (\xi -a) m + \sigma z +h \nonumber \\
& & + \alpha \left[ \frac{J^2}{1-JU} 
+ (\tilde{J^2}-J^2) \right] U Y(z;\xi ) \biggr). \nonumber \\
\label{ap:nonlinear.Y}
& &
\end{eqnarray}
Finally, 
the equivalence between the nonlinear synapse 
and the additive synaptic noise is obtained as follows, 
\begin{eqnarray}
\Delta_A^2 &=& \alpha (1-a^2)^2 \biggl( \frac{\tilde{J^2}}{J^2}-1 \biggr), 
\label{ap:nonlinear2add} \\
J &=& \int Dx \; x f(x) \\
\tilde{J^2} &=& \int Dx f(x)^2
\end{eqnarray}
by comparing 
Eqs.(\ref{ap:nonlinear.sigma}) and (\ref{ap:nonlinear.Y}) 
to Eqs.(\ref{ap:sigma.additive_noise}) 
and (\ref{ap:effective_response_function}). 
Substituting Eq.(\ref{ap:nonlinear2add}) into Eq.(\ref{ap:mul2add}), 
we obtain the equivalence of Eq.(\ref{5eq:nonlinear.to.mul}).

\section{Asymptote for large multiplicative synaptic noise}
\label{appendix.critical_noise}

Derivations of the asymptote of storage capacity 
in a large multiplicative synaptic noise $\Delta_M$ is given here. 

In Eqs.(\ref{eq:order_parameter_equation_m})
-(\ref{eq:order_parameter_equation_U}), 
let $a=0$, $J=1$, and $F(x)=\sgn (x)$, 
the order-parameter equations become 
\begin{eqnarray}
m &=& \erf \biggl( \frac{m}{\sqrt{2} \sigma} \biggr) , \label{ap:m} \\
q &=& 1, \\
U &=& \frac 1\sigma \sqrt{\frac{2}{\pi}} 
      \exp \biggl( -\frac{m^2}{2\sigma^2} \biggr) , \label{ap:U}
\end{eqnarray}
the threshold becomes $h=0$, 
the effective response function of 
Eq.(\ref{eq:effevtive_response_function}) 
and the variance of the noise become 
\begin{eqnarray}
Y(z;\xi ) &=& {\rm sgn}(\xi m + \sigma z), \\
\sigma^2 &=& \frac {\alpha}{(1-U)^2} + \Delta_A^2, \label{ap:sigma2}
\end{eqnarray}
respectively, 
where the error function ${\rm erf}(x)$ is defined as 
\begin{equation}
{\rm erf}(x) = \frac2{\sqrt{\pi}} \int_0^x e^{-u^2} du. 
\end{equation}
The slope of the r.h.s. of Eq.(\ref{ap:m}) is given by 
\begin{equation}
\frac {d}{dm} \erf \biggl( \frac{m}{\sqrt{2} \sigma} \biggr)
= \frac 1\sigma \sqrt{\frac 2 \pi} 
  \exp \biggl( -\frac {m^2}{2\sigma^2} \biggr) . \label{ap:erf_slope}
\end{equation}
Equation (\ref{ap:m}) has nontrivial solutions $m\ne 0$ 
within the range where the slope of the r.h.s. of Eq.(\ref{ap:erf_slope}) at $m=0$ is greater than 1.
Therefore, the critical value of the noise $\sigma_c^2$ is given by 
\begin{equation}
\sigma_c^2 = 2/\pi.
\end{equation}
This shows that a retrieval phase exists only for $\sigma<\sigma_c$. 
We define the parameter $\tau (<1)$ defined as 
\begin{equation}
\tau=\frac \sigma{\sigma_c}, 
\end{equation}
to solve for $m$ as a function of $\sigma$ 
in the vicinity of this critical value $\sigma_c$. 
The critical value of the additive synaptic noise 
is discussed in the case of $\tau \simeq 1$. 
The overlap $m$ shows the first order phase transition 
when $\Delta_A$ is small, 
but it is regarded as the second order phase transition at large $\Delta_A$ region. 
The overlap becomes $m \ll 1$ when $\tau \simeq 1$ 
and $\Delta_A$ is sufficiently large, 
therefore the nontrivial solution of $m$ is given as 
\begin{equation}
m
\simeq
\frac m\tau - \frac {m^3}{6\sigma^2\tau} + O(m^4)
=
\sigma_c \tau \sqrt{6(1-\tau )}, \label{ap:nontrivial_m}
\end{equation}
by Taylor expansion including terms up to the third order. 
Substituting Eq.(\ref{ap:nontrivial_m}) into Eq.(\ref{ap:U}), $U$ becomes 
\begin{equation}
U
\simeq
\frac 1\tau \biggl(  1-\frac {m^2}{2\sigma^2} \biggr) +O(m^4) 
=
3 -2\tau^{-1} . \label{ap:U_at_small_m}
\end{equation}
From Eq.(\ref{5eq:add.to.mul}),
the variance of the multiplicative synaptic noise $\Delta_M^2$ 
is related to the variance of the additive synaptic noise $\Delta_A^2$ as 
\begin{equation}
\Delta_A^2 = \alpha \Delta_M^2 , \label{ap:a2m_at_a=0}
\end{equation}
when bias $a=0$. 
Therefore, substituting Eqs.(\ref{ap:U_at_small_m}) and (\ref{ap:a2m_at_a=0}) into Eq.(\ref{ap:sigma2}), 
the variance of the noise $\sigma$ is given as 
\begin{equation}
\sigma^2=\frac{\alpha \tau^2}{4(1-\tau )^2}+\alpha \Delta_M^2. 
\label{ap:sigma2_aprx}
\end{equation}
The loading rate $\alpha$ becomes 
\begin{equation}
\alpha = \frac 8\pi \cdot \frac{\tau^2(1 - \tau)^2}{\tau^2 + 4 \Delta_M^2 (1 - \tau^2)}.
\label{ap:small_m_op_eq}
\end{equation}
When the overlap is small enough, i.e., $m \ll 1$, 
the order-parameter equations of Eqs. (\ref{ap:m})-(\ref{ap:sigma2}) reduce to Eq. (\ref{ap:small_m_op_eq}). 
Solving Eq.(\ref{ap:small_m_op_eq}) for the fixed value of $\alpha$ and $\Delta_M$, we obtain the parameter $\tau$. 
Substituting $\tau$ into Eq. (\ref{ap:nontrivial_m}), we can obtain the overlap $m$ for given $\alpha$ and $\Delta_M$. 
It is easily confirmed that the $\tau$ increases with $\alpha$ for the fixed value of $\Delta_M$. 
This means that the maximal value of $\tau$ which holds Eq. (\ref{ap:small_m_op_eq}) 
corresponds to the maximum value of $\alpha$, that is storage capacity $\alpha_c$. 
The critical value $\tau_c$ is equal to the value which maximizes the loading rate of Eq.(\ref{ap:small_m_op_eq}) 
and becomes 
\begin{equation}
\tau_c = \frac{(2\Delta_M)^{2/3}}{1+(2\Delta_M)^{2/3}} \simeq 1-(2\Delta_M)^{-2/3}, \label{ap:tau_c}
\end{equation}
in a large $\Delta_M$ limit.
Therefore, substituting Eq. (\ref{ap:tau_c}) into Eq.(\ref{ap:small_m_op_eq}), 
we obtain Eq.(\ref{5eq:mul.asymptote}) as follows: 
\begin{equation}
\alpha_c =\frac 2{\pi \Delta_M^2}. 
\end{equation}
\end{document}